\documentclass[conference]{IEEEtran}
\IEEEoverridecommandlockouts
\usepackage{cite}

\usepackage{algorithm}
\usepackage{algorithmic}
\usepackage{graphicx}
\usepackage{subcaption}
\usepackage{tabularx}

\usepackage{amsmath,amssymb,amsfonts}
\usepackage{algorithmic}
\usepackage{graphicx}
\usepackage{textcomp}
\usepackage{xcolor}
\def\BibTeX{{\rm B\kern-.05em{\sc i\kern-.025em b}\kern-.08em
    T\kern-.1667em\lower.7ex\hbox{E}\kern-.125emX}}
\begin{document}

\title{ProbSelect: Stochastic Client Selection for GPU-Accelerated Compute Devices in the 3D Continuum\\
\thanks{This research is funded by the EU's Horizon Europe Research and Innovation Program as part of the NexaSphere project (GA No. 101192912).}
}

\author{\IEEEauthorblockN{Andrija Stanisic}
\IEEEauthorblockA{\textit{Distributed Systems Group} \\
\textit{TU Wien}\\
a.stanisic@dsg.tuwien.ac.at}
\and
\IEEEauthorblockN{Stefan Nastic}
\IEEEauthorblockA{\textit{Distributed Systems Group} \\
\textit{TU Wien}\\
s.nastic@dsg.tuwien.ac.at}
}

\maketitle

\begin{abstract}

Integration of edge, cloud and space devices into a unified 3D continuum imposes significant challenges for client selection in federated learning systems. Traditional approaches rely on continuous monitoring and historical data collection, which becomes impractical in dynamic environments where satellites and mobile devices frequently change operational conditions. Furthermore, existing solutions primarily consider CPU-based computation, failing to capture complex characteristics of GPU-accelerated training that is prevalent across the 3D continuum. This paper introduces ProbSelect, a novel approach utilizing analytical modeling and probabilistic forecasting for client selection on GPU-accelerated devices, without requiring historical data or continuous monitoring. We model client selection within user-defined SLOs. Extensive evaluation across diverse GPU architectures and workloads demonstrates that ProbSelect improves SLO compliance by 13.77\% on average while achieving 72.5\% computational waste reduction compared to baseline approaches.


\end{abstract}

\begin{IEEEkeywords}
3D Compute Continuum, federated learning, client selection, Compound AI
\end{IEEEkeywords}

\section{Introduction}

The 3D Compute Continuum marks an evolution of the traditional Edge–Cloud model into an integrated Edge–Cloud–Space paradigm, seamlessly uniting data centers, edge, terrestrial, aerial, and space-based devices into a cohesive computing ecosystem \cite{pusztai2024hyperdriveschedulingserverlessfunctions}. This enables dynamic computation distribution across all layers, adapting to varying latency, bandwidth, and privacy requirements. As computational demands increase, devices at all layers are equipped with powerful hardware accelerators such as Graphics Processing Units (GPUs), with modern satellites now participating in complex tasks like real-time wildfire detection \cite{pusztai2024hyperdriveschedulingserverlessfunctions}.
Since these nodes become more capable, they generate increasingly large volumes of sensitive data that cannot be easily centralized due to governance rules, privacy constraints and its sheer scale. To address these constraints while leveraging computational capabilities, federated learning enables collaborative model training without sharing raw data \cite{FL-base}. 
Nevertheless, federated learning deployments often involve hundreds of thousands of heterogeneous devices.
Rather than including all available devices in each training round, intelligent participant selection can significantly improve system performance \cite{lai2021oortefficientfederatedlearning}. 


However, participant selection in heterogeneous environments introduces several challenges. The diverse nature of devices across the 3D Compute Continuum creates substantial variations in compute power, memory availability, and network conditions. One of the core challenges that emerges in client selection is the \textit{straggler problem} \cite{reisizadeh2020stragglerresilientfederatedlearningleveraging}.
Stragglers are devices that are significantly slower than others during training rounds. When slower devices delay task completion, they can bottleneck the aggregation step and tremendously increase training latency \cite{9084352}. Many existing solutions set fixed training deadlines and accept only updates arriving within that window \cite{survey}. Introducing training deadlines prevents fast devices from idling but creates inefficiency when most selected participants miss deadlines. As their updates are not being acknowledged, they discard valuable computational and network resources.

To address the straggler problem, many approaches utilize analytical models to predict device performance and intelligently select clients that are more likely to meet predefined deadlines. Unfortunately, most current solutions focus primarily on CPU-based computational models \cite{8761315,10001569,9685077}, 
neglecting the specific characteristics of GPU-accelerated training, such as parallel execution patterns, memory hierarchies, and efficiency variations that fundamentally determine GPU training performance. This leads to inaccuracies in performance predictions as devices across 3D Compute Continuum increasingly adopt GPU accelerators for machine learning workloads. Additionally, many existing analytical approaches require historical data collection or extensive benchmarking procedures to function effectively \cite{9052206}. These requirements introduce resource overhead and computational complexity, making them unsuitable for large-scale heterogeneous systems. This challenge is particularly pronounced in the 3D Compute Continuum, where satellites traveling at high orbital velocities rapidly move in and out of communication range \cite{pusztai2024hyperdriveschedulingserverlessfunctions}
, making continuous monitoring extremely difficult. 



Our work addresses these challenges by developing a novel approach that combines analytical modeling with probabilistic client selection to address the straggler problem in GPU-accelerated federated learning. The proposed solution eliminates the need for historical data collection or continuous monitoring, making it compatible with dynamic environments such as the 3D continuum. 
The main contributions of this work include:

\begin{itemize}

    \item \textbf{ALM}: \textit{A novel Analytical Latency Model for GPU-Accelerated Training} that estimates the training latency on GPU-enabled edge devices (Section~\ref{section:models}). ALM has a primary focus on accurately modeling GPU training time through analytical characterization of parallel execution patterns, memory hierarchies, and hardware-specific efficiency factors. Evaluation across diverse GPU architectures and neural network models demonstrates ALM prediction accuracy with an overall Mean Absolute Percentage Error (MAPE) below 5\% and individual prediction errors ranging from 0.5\% to 12.8\%.

    \item \textbf{ProbSelect}: \textit{A novel Probability-Based Client Selection Algorithm} that leverages our ALM model to identify devices with high probability of meeting predefined service level objectives (SLOs) (Section~\ref{section:ProbSelect}). ProbSelect utilizes ALM alongside user-defined deadlines and probability thresholds to make intelligent participant selection decisions, thereby reducing wasted computation and improving overall system efficiency in federated training scenarios. Our experimental evaluation (Section~\ref{section:experiments}) shows that ProbSelect improves SLO compliance by 13.77\% on average while achieving a reduction in computational waste of up to 72.5\% compared to baseline approaches.
\end{itemize}

\subsection{Limitations}




The primary objective of our work centers on deadline-aware client selection rather than end-to-end federated learning optimization. We define computational waste as clients selected but failing to complete training before specified deadlines, and optimize specifically for meeting those deadlines, which are defined by SLO compliance metrics. Consequently, this work does not evaluate nor consider end-to-end training performance indicators such as model convergence speed, convergence time, or global model accuracy. Although these metrics are important for comprehensive federated learning evaluation, they fall outside our research scope of energy- and SLO-aware resource optimization.


\section{System Assumptions, Models and Problem Formulation}
\label{section:models}


\subsection{Assumptions}

Let $\mathcal{I}$ denote the set of edge devices participating in the federated learning process. We consider a scenario where devices are distributed within a base station's coverage area and communicate with a centralized federated learning server through wireless links. Each device $i \in \mathcal{I}$ is equipped with a dedicated GPU exclusively allocated to federated learning tasks during training rounds. In each round, we randomly select $\mathcal{K} \subseteq \mathcal{I}$ devices to undergo the selection process. Furthermore, we assume that each device $i \in \mathcal{K}$, generates device profile $\mathcal{E}_i$ and distributes it to the centralized server. Consequently, centralized server will store complete set of device profiles defined as:
\begin{equation}
   \mathcal{E} = \{\mathcal{E}_i : i \in \mathcal{K} \}
\end{equation}

Each device profile $\mathcal{E}_i$ contains GPU hardware specifications, current network upload and download bandwidths, and local dataset size. These profiles are collected and distributed to centralized server prior to each selection process and serve as key enablers for the informed client selection. 

Regarding network infrastructure, we consider a Time Division Duplex (TDD) wireless communication system where devices share available bandwidth sequentially \cite{amiri2020convergenceupdateawaredevice}. We assume stable network conditions during training rounds and model communication latency through device-specific upload and download bandwidth rates, following established practices in federated learning literature \cite{ 8761315}. 
Finally, we consider server-side weight aggregation and broadcasting latencies negligible compared to device-level computation and communication.

\subsection{Analytical Latency Model}

To systematically account for the primary sources of latency in federated training rounds, we propose the Analytical Latency Model (ALM) for GPU-accelerated edge devices. ALM decomposes the total training latency into three distinct phases, \textit{download}, \textit{compute}, and \textit{upload}, each influenced by both network conditions and device-specific GPU capabilities. The total latency for device $i$ is given by:
\begin{equation}
\tau_i = \tau^{\text{download}}_i + \tau^{\text{compute}}_i + \tau^{\text{upload}}_i
\label{eq:latency}
\end{equation}

Each term in Equation~\ref{eq:latency} represents a specific stage of the training process, contributing to the overall latency.

\subsubsection{Download Latency}

Download latency is calculated based on the available network bandwidth at the client device:
\begin{equation}
\tau^{\text{download}}_i = \frac{m_s}{a^{\text{download}}_i}
\label{download_latency_model}
\end{equation}

where \(m_s\) denotes the model size in bytes, which is known to the server, and \(a^{\text{download}}_i\) is the available download bandwidth of device \(i\).

\subsubsection{Upload Latency}

Upload latency is computed similarly:
\begin{equation}
\tau^{\text{upload}}_i = \frac{m_s}{a^{\text{upload}}_i}
\label{upload_latency_model}
\end{equation}

where \(a^{\text{upload}}_i\) denotes the available upload bandwidth of device \(i\).

\subsubsection{Computational Latency}

The computational latency represents the dominant component of the total training time in GPU-accelerated federated learning \cite{Pan_2024_BMVC}. We propose an analytical model that decomposes this latency into data transfer and processing components. The total computational latency is expressed as:
\begin{equation}
    \tau_i^{compute} = \tau_i^{model \ load} + \theta log_2(\dfrac{1}{\epsilon}) \tau_i^{train}
    \label{comp_base}
\end{equation}

where the first term captures initial loading of model weights, and the second term models iterative batch processing over multiple local training epochs.

The model loading component represents the time required to transfer the global model parameters from system memory to GPU memory via the PCI Express interface:
\begin{equation}
    \tau_i^{model \ load} = \dfrac{m_s}{PCIb}
    \label{model_transfer}
\end{equation}

here $PCI_b$ represents device-specific PCI Express bandwidth. The latency introduced during this loading process occurs once at the start of each global training round, when the global model is transferred to the GPU for local updates.

The number of local training epochs required to achieve desired model accuracy $\epsilon$ is determined by $\theta \log_2(\frac{1}{\epsilon})$, where $\theta$ is a convergence parameter derived from gradient descent theory for strongly convex functions \cite{yang2020delayminimizationfederatedlearning}. Equation \ref{batch_comp} models latency introduced at device $i$ while executing one training epoch.
\begin{equation}
    \tau_i^{train} = \dfrac{|D_i|}{|b_s|} \times (\dfrac{G \times |b_s|}{\omega_i f_i p_i \eta_i} + \dfrac{b_s}{PCIb})
    \label{batch_comp}
\end{equation}

Here, $|D_i|$ defines the number of local training samples on device $i$. Term $|b_s|$ quantifies the number of samples per training batch. Consequently, their ratio $\frac{|D_i|}{|b_s|}$ expresses the number of batches to be processed. Equation \ref{batch_comp} comprises two distinct components that capture the fundamental bottlenecks in GPU-accelerated training. The first term within the parentheses quantifies the computational workload, where $G$ represents the total number of floating-point operations (FLOPs) required for executing forward propagation, backward propagation, and parameter optimization for a single training sample. The GPU's computational capacity is modeled as $\omega_i f_i p_i$, where $\omega_i$ denotes the architecture-specific number of operations per clock cycle, $f_i$ represents the GPU boost clock frequency, and $p_i$ indicates the number of GPU cores. The efficiency factor $\eta_i$ represents the proportion of theoretical FLOPs effectively used during training, allowing for hardware and software specific constraints that prevent full theoretical utilization. 


The second term, $\frac{b_s}{PCI_b}$, accounts for the latency incurred when transferring training samples from system memory to GPU memory during batch processing, where $b_s$ represents the total batch size in bytes. Substituting Equations \ref{model_transfer} and \ref{batch_comp} into Equation \ref{comp_base} gives:
\begin{equation}
    \tau_i^{compute} = \dfrac{m_s}{PCIb} + \theta log_2(\dfrac{1}{\epsilon})\dfrac{|D_i|}{|b_s|}(\dfrac{G \times |b_s|}{\omega_i f_i p_i \eta_i} + \dfrac{b_s}{PCIb})
    \label{almost main}
\end{equation}

To further simplify Equation \ref{almost main}, we utilize the relationship $b_s = s_s \times |b_s|$, where $s_s$ represents input sample size in bytes, yielding the final computational latency formulation:
\begin{equation}
    \tau_i^{compute} = \dfrac{m_s}{PCIb} + \theta log_2(\dfrac{1}{\epsilon})|D_i|(\dfrac{G}{\omega_i f_i p_i \eta_i} + \dfrac{s_s}{PCIb})
    \label{main}
\end{equation}

A critical challenge in applying the proposed analytical model is determining the efficiency factor $\eta_i$. This parameter cannot be derived solely from device profiles, as it depends on complex workload-hardware interactions. Determining $\eta_i$ typically introduces computational overhead and system complexity through runtime monitoring or benchmarking procedures. As part of our evaluation, we extract efficiency factors by benchmarking real training workloads on target GPU devices and use them only as a baseline to validate ALM. However, our proposed solution in Section \ref{section:ProbSelect} addresses this challenge by eliminating the dependency on exact efficiency factor values, thus removing the need for historical data collection or monitoring.

\subsection{Client Selection Problem Formulation}


Given the complete set of device profiles $\mathcal{E}$ and the ALM model, we formulate the client selection problem. The user defines two key parameters: $\tau^{\text{slo}}$, the maximum allowable duration for one global training round, and $p^{\text{slo}}$, the probability threshold used as a selection criterion. 

For device $i \in \mathcal{K}$ to be selected, the following constraint must be satisfied:
\begin{equation}
\label{slo_constraint}
p_i \geq p^{\text{slo}}
\end{equation}

where $p^{\text{slo}}$ represents the user-defined probability threshold, and $p_i$ denotes the probability of device $i$ completing training before $\tau^{\text{slo}}$. The total training latency $\tau_i$ for device $i$ is modeled using ALM. We can formally express the probability of device $i$ finishing training before the deadline as:
\begin{equation}
\label{probability_definition}
p_i = P(\tau_i \leq \tau^{\text{slo}})
\end{equation}

Finally, let $\mathcal{S}$ denote the set of selected clients. Based on the probabilistic constraint $p_i$, we define the client selection strategy by selecting all devices that meet the probability threshold $p^{slo}$. Equation \ref{selection_strategy} formally defines our selection strategy.
\begin{equation}
\label{selection_strategy}
\mathcal{S} = \{i \in \mathcal{K} : p_i \geq p^{\text{slo}}\}
\end{equation}



\section{ProbSelect: Client Selection Algorithm}
\label{section:ProbSelect}

We propose ProbSelect, a probability-based client selection algorithm that utilizes ALM to identify devices with high probability of meeting predefined SLO deadlines. ProbSelect leverages efficiency thresholds and probabilistic modeling to make intelligent participant selection.

\subsection{GPU Efficiency Threshold Calculation}

To determine the minimum GPU efficiency required for deadline compliance, we derive the efficiency threshold for each device by rearranging proposed latency model. Starting from the deadline constraint $\tau_i \leq \tau^{slo}$, we consider the boundary case where $\tau_i = \tau^{slo}$ to establish the exact efficiency threshold required for device $i$ to complete training precisely at the deadline:
\begin{equation}
    \tau_{i}^{download} + \tau_i^{compute} + \tau_i^{upload} = \tau^{slo}
\end{equation}

We isolate the computational latency by substituting our download and upload models, yielding the available time for computation:
\begin{equation}
    \tau' = \tau^{slo} - \left(\frac{m_s}{a_i^{download}} + \frac{m_s}{a_i^{upload}}\right)
    \label{eq:available_comp_time}
\end{equation}
Substituting the computational latency model from Equation \ref{main} and rearranging to isolate the efficiency dependent term gives:
$$ \theta log_2(\dfrac{1}{\epsilon})|D_i|(\dfrac{G}{\omega_i f_i p_i \eta_i^{th}}) = \tau^{'} - (\dfrac{m_s}{PCIb} + \theta log_2(\dfrac{1}{\epsilon})|D_i|\dfrac{s_s} {PCIb})$$

We define the net computing time $\tau''$ as the time exclusively available for GPU computation after accounting for all data and network overheads:
\begin{equation}
    \tau'' = \tau' - \left(\frac{m_s}{PCIb} + \theta \log_2\left(\frac{1}{\epsilon}\right)|D_i|\frac{s_s}{PCIb}\right)
\end{equation}

Finally, the threshold efficiency is given by the following relation:
\begin{equation}
    \eta_i^{th} = \theta \log_2\left(\dfrac{1}{\epsilon}\right)|D_i|\left(\dfrac{G}{\omega_i f_i p_i \tau^{''}}\right)
    \label{nth}
\end{equation}

Efficiency threshold $\eta_i^{th}$ represents the minimum percentage of theoretical GPU FLOPs that device $i$ must achieve to complete the entire training process exactly at the SLO deadline. Any efficiency value exceeding this threshold guarantees deadline compliance:
\begin{equation}
    \forall \eta_i \geq \eta^{th}_i \Rightarrow \tau_i \leq \tau^{slo}
\end{equation}

\subsection{Probabilistic Client Selection}



We assume that for a given GPU and workload, the FLOP utilization $\eta_i$ follows a normal distribution $\mathcal{N}(\mu_i, \sigma_i^2)$. Here, $\mu_i$ represents the expected FLOP utilization for the specific model and dataset on device $i$, and $\sigma_i^2$ captures the efficiency variations across different training iterations due to transient factors such as memory access patterns and runtime optimizations. To the best of our knowledge, no current research quantifies GPU FLOP efficiency distributions, so we adopt this assumption following the well-established practice of modeling real-world phenomena with normal distributions \cite{book}. With this assumption, we can calculate the probability that device $i$ will achieve efficiency higher than the required threshold, thus completing training before the SLO deadline $\tau^{slo}$:
\begin{equation}
p_i = P(\eta_i \geq \eta_i^{th}) = 1 - \Phi\left(\frac{\eta_i^{th} - \mu_i}{\sigma_i}\right)
\label{eq:probability_calculation}
\end{equation}

where $\Phi(\cdot)$ denotes the cumulative distribution function (CDF) of the standard normal distribution. Using these probability values, we can now apply the client selection criterion defined in our problem formulation (Equation 12) to compose the set $S \subseteq \mathcal{K}$ of selected devices for participation in the next training round. 

\begin{algorithm}
\caption{\textit{ProbSelect}}
\label{alg:ProbSelect_detailed}
\begin{algorithmic}[1]
\REQUIRE $\mathcal{E}$, $m_s$, $s_s$, $G$, $\tau^{slo}$ $p^{slo}$, $\mu_i, \sigma_i$.
\ENSURE Set of selected devices $\mathcal{S}$
\FOR{$i \in K$}
    \STATE Calculate $\eta_i^{th}$ using Eq. \ref{nth}
    \STATE Calculate $p_i$ using Eq. \ref{eq:probability_calculation}
    \STATE $S = \emptyset$
    \IF{$p_i \geq p^{slo}$}
        \STATE $\mathcal{S} = \mathcal{S} \cup i$ 
    \ENDIF
\ENDFOR
\RETURN $\mathcal{S}$
\end{algorithmic}
\end{algorithm}

ProbSelect, summarized in Algorithm~\ref{alg:ProbSelect_detailed}, operates in three main steps. First, it computes efficiency thresholds for all devices $i \in \mathcal{K}$ using Equation~\ref{nth}, determining the minimum GPU FLOP efficiency required for deadline compliance. Second, it calculates the probability of each device achieving GPU efficiency above the computed threshold using Equation~\ref{eq:probability_calculation}, thereby transforming deterministic requirements into probabilistic assessments. Finally, it selects devices satisfying the constraint in Equation~\ref{selection_strategy}, retaining only those whose compliance probability exceeds $p^{slo}$.

\section{Evaluation}
\label{section:experiments}

\subsection{Analytical Latency Model Evaluation}


To evaluate the standalone performance of ALM, we conduct a series of experiments on different GPUs. Access to these computational resources was facilitated through the high-performance computing (HPC) cluster at TU Wien.

\subsubsection{Experiment setup}

The GPUs we used are from NVIDIA and include: RTX 4090, Tesla V100, A100, A40, and Tesla T4. This heterogeneous selection strategically spans the computational spectrum of the 3D continuum: from resource-constrained edge devices (Tesla T4) to high-performance cloud infrastructure (A100), ensuring our approach is applicable across edge, cloud, and space environments where similar GPU architectures are increasingly deployed. We utilized the \textit{pynvml} \cite{pynvml} library to record GPU specifications directly from hardware, as summarized in Table \ref{tab:gpu_specs}.

\begin{table}[h]
\centering
\caption{GPU Specifications Used in ALM Evaluation.}
\resizebox{0.49\textwidth}{!}{
\begin{tabular}{|l|l|c|c|c|c|}
\hline
\textbf{GPU} & \textbf{Architecture} & $\boldsymbol{\omega_i}$ & $\boldsymbol{p_i}$ & $\boldsymbol{f_i}$ (\textbf{MHz}) & \textbf{PCIe} (\textbf{GB/s}) \\
\hline
RTX 4090    & Ada Lovelace & 2 & 16384 & 2520 & 31.5 \\
\hline
Tesla V100  & Volta        & 2 & 5120  & 1380 & 15.75 \\
\hline
A100   & Ampere       & 2 & 6912  & 1410 & 31.5 \\
\hline
A40         & Ampere       & 2 & 10752 & 1740 & 31.5 \\
\hline
Tesla T4    & Turing       & 2 & 2560  & 1590 & 15.75 \\
\hline
\end{tabular}
}
\label{tab:gpu_specs}
\end{table}


To complement this diverse hardware setup, our workload comprises model and dataset combinations reflecting realistic deployment scenarios across the 3D continuum. We focus on convolutional neural networks to support real-life scenarios where satellites perform image classification tasks. The models used are ResNet-50 \cite{he2016deep}, AlexNet \cite{krizhevsky2012imagenet}, and MobileNetV2 \cite{sandler2018mobilenetv2}, trained on CIFAR-100,
TinyImageNet 
and CIFAR-10
datasets respectively. This diverse combination enables us to evaluate ALM performance across heterogeneous workloads. To extract model characteristics required for ALM evaluation, we utilize library \textit{torchinfo} \cite{torchinfo}, which provides measurements of FLOPs needed to process single sample $G$, model size $m_s$, and input sample size $s_s$. Table \ref{tab:workloads} summarizes the experimental workloads used in our evaluation:

\begin{table}[h]
\centering
\caption{Experimental Workloads for ALM Evaluation.}
\resizebox{0.49\textwidth}{!}{%
\begin{tabular}{|l|c|l|c|c|c|}
\hline
\textbf{Model} & $\boldsymbol{m_s}$ (\textbf{MB}) & \textbf{Dataset} & $\boldsymbol{|D_i|}$ & $\boldsymbol{G}$ (\textbf{GFLOPs}) & $\boldsymbol{s_s}$ (\textbf{MB}) \\
\hline
ResNet-50 & 97.49 & CIFAR-100 & 50.000 & 24.53 & 0.6 \\
\hline
AlexNet & 233.08 & Tiny-ImageNet & 100.000 & 4.28 & 0.6 \\
\hline
MobileNetV2 & 13.37 & CIFAR-10 & 50.000 & 1.80 & 0.6 \\
\hline
\end{tabular}
}
\label{tab:workloads}
\end{table}

Our experiments are organized to first extract the FLOPs efficiency factor $\eta_i$ for a given workload on each GPU. We then use this efficiency to estimate performance of subsequent training rounds. To calculate the actual achieved efficiency of a given workload on a specific GPU, we measure the actual training time and utilize Equation \ref{nth}. Here we set $\theta \log_2(\frac{1}{\epsilon}) = 1$, to calculate efficiency for one round, and we substitute the net computing time $\tau''$ with the actual measured training time.

Using the extracted efficiency factors, we validate ALM's predictive capability for 10-epoch estimation. We focus on this range because our optimization target is individual global training rounds, where each selected device typically performs small number of local iterations before uploading model updates \cite{8761315, FL-base}.

\begin{figure}[h!]
   \centering
   \includegraphics[width=0.47\textwidth]{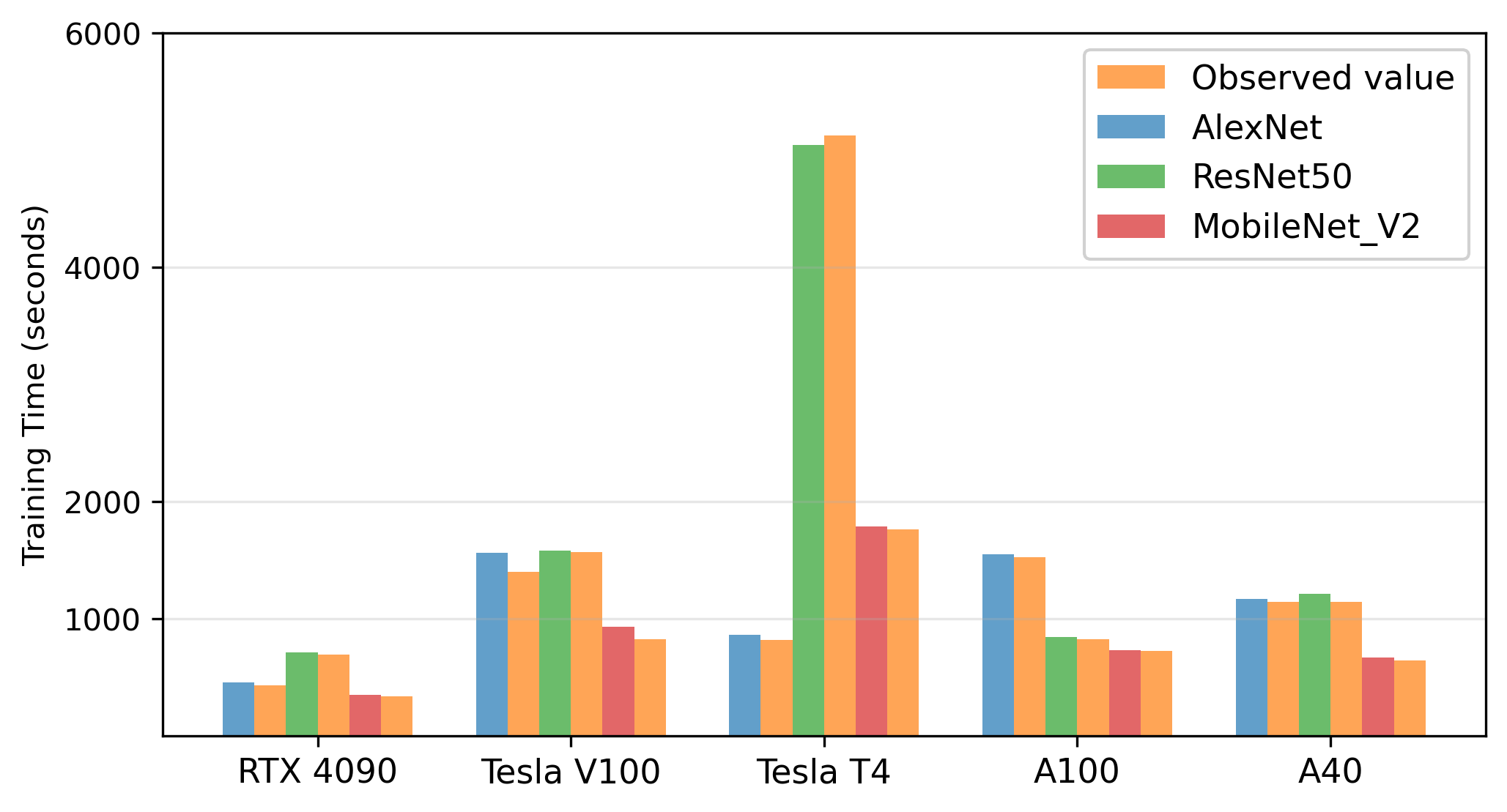}
   \caption{Estimated vs. measured training times across GPU architectures for ten training epochs.}
   \label{fig:training_comparison}
\end{figure}
\subsubsection{Experiment results}
Figure \ref{fig:training_comparison} shows the comparison between actual measured training times and ALM-predicted training times for ten-epoch scenarios across different GPU architectures and workloads. The results illustrate ALM's accuracy in ten-epoch predictions, achieving an overall MAPE of 4.2\%. These results validate that our analytical formulas correctly model GPU training behavior, providing confidence that the same mathematical foundations can be reliably used in ProbSelect's efficiency threshold calculations. ALM's efficiency factor extraction successfully captures the underlying computational characteristics across different hardware configurations. While individual predictions range from 0.5\% to 12.8\% APE across different hardware-workload combinations, this experimental validation establishes the reliability of the analytical foundations that enable ProbSelect's probabilistic client selection methodology.

\subsection{Client Selection Algorithm Evaluation}
For the purpose of evaluating ProbSelect we compare it to the FedLim \cite{8761315}, an algorithm that employs random client selection while incorporating deadline awareness.

\subsubsection{Experiment setup}

We observe an environment with $\mathcal{I} = 1000$ devices, where devices have one of the 5 aforementioned GPUs as their main computational resource. Each GPU is randomly distributed and present in 20\% of total participating devices. We utilize real hardware measurements for each workload, defined in Table~\ref{tab:workloads}, and set a variance in training iterations to 20\% to achieve a simulated environment based on real-life measurements. For the FLOP utilization distribution, we empirically derive and set $\mu_i = 0.5$ and $\sigma_i = 0.25$. The mean $\mu_i = 0.5$ represents realistic average utilization given hardware and software constraints that prevent full theoretical FLOP achievement. The standard deviation $\sigma_i = 0.25$ enables gradual client selection decisions in our probabilistic framework, creating smooth probability transitions rather than harsh selection cutoffs. Devices requiring 30\% utilization achieve 78.8\% selection probability while those requiring 70\% utilization maintain 21.2\% selection probability. In contrast, smaller variance (e.g., $\sigma_i = 0.1$) would create rigid cutoffs where devices requiring 30\% utilization achieve 97.7\% selection probability while those requiring 70\% utilization drop to only 2.3\% selection probability. This gradual selection behavior is particularly beneficial in dynamic 3D continuum environments where rigid selection thresholds could lead to suboptimal resource utilization under constantly changing operational conditions.


We conduct experiments over one hundred global training epochs. In each training epoch, $\mathcal{K} = 100$ devices are randomly selected for client selection. For each workload, we empirically identify the probability and deadline SLO values, summarized in Table~\ref{tab:slo_params}.
\begin{table}[h!]
\centering
\caption{SLO parameters for each workload.}
\label{tab:slo_params}
\begin{tabular}{|l|c|c|}
\hline
\textbf{Model} & $\boldsymbol{\tau^{slo}}$ (\textbf{s}) & $\boldsymbol{p^{slo}}$ (\textbf{\%}) \\
\hline
ResNet-50 & 50 & 90 \\
\hline
AlexNet & 125 & 90 \\
\hline
MobileNetV2 & 100 & 90 \\
\hline
\end{tabular}
\end{table}


Regarding the network, we model asymmetric bandwidth profiles with upload rates ranging from 83 to 181 Mbps and download rates from 650 to 830 Mbps, based on real-world 5G standalone network measurements \cite{LACKNER20221132}. Before each round, devices are assigned random values from these ranges for upload and download network rates. 
\begin{figure}[h!]
    \centering
    \begin{subfigure}[t]{0.24\textwidth}
        \includegraphics[width=\linewidth]{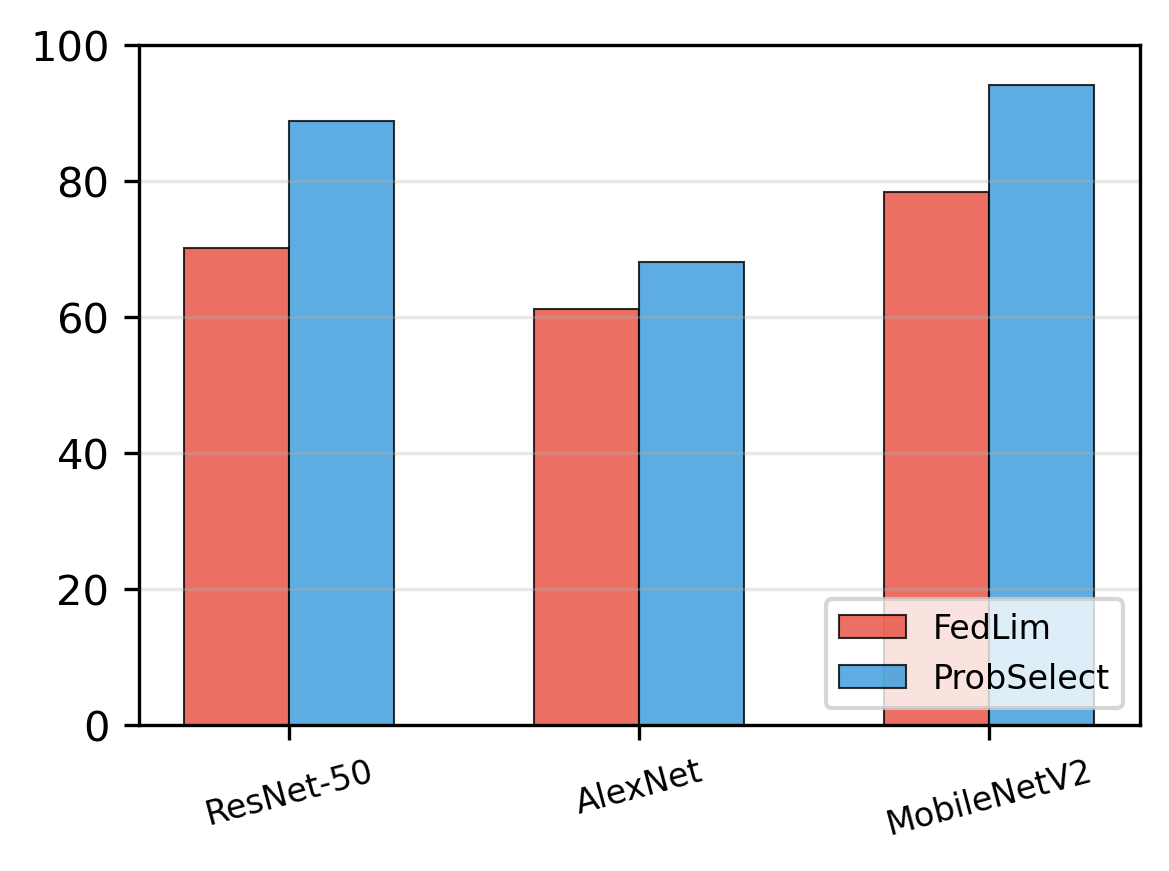}
        \caption{SLO compliance rate.}
        \label{fig:slo_compliance}
    \end{subfigure}
    \begin{subfigure}[t]{0.24\textwidth}
        \includegraphics[width=\linewidth]{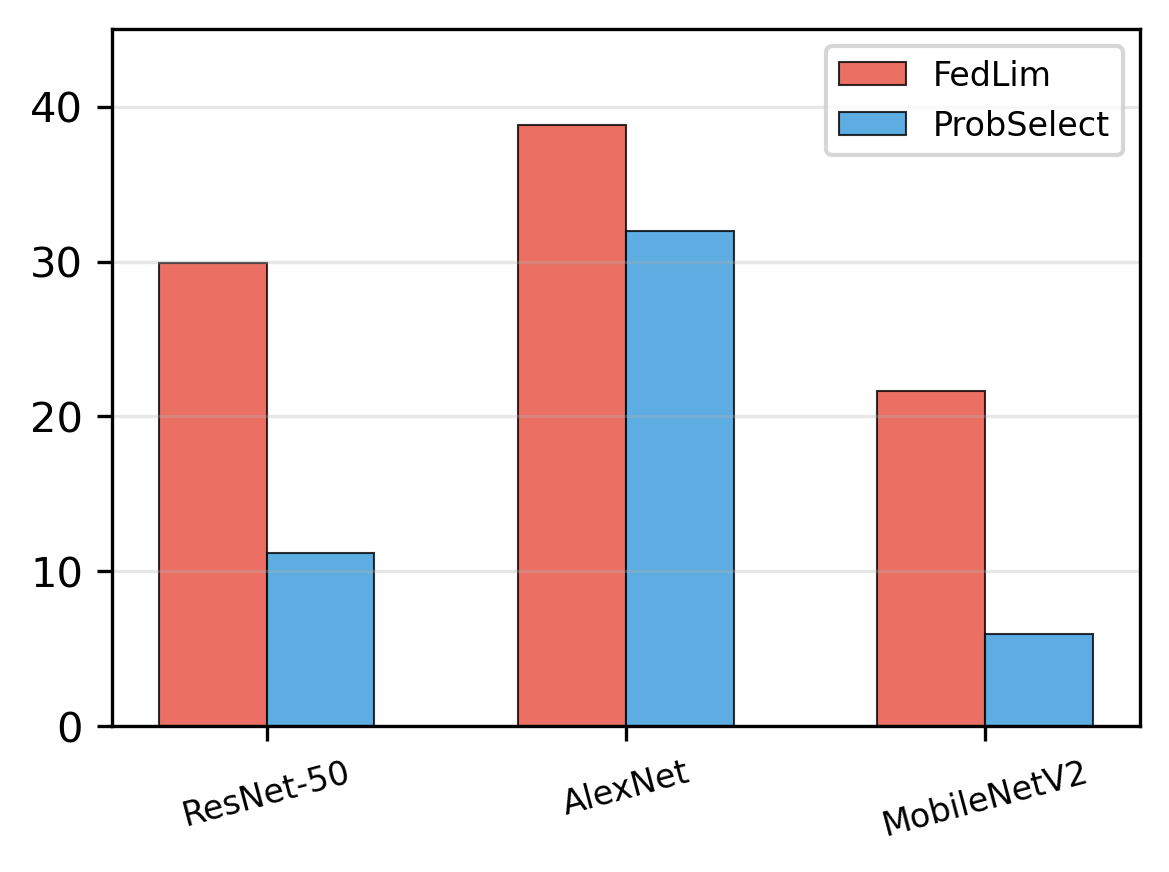}
        \caption{Computational waste rate.}
        \label{fig:waste_reduction}
    \end{subfigure}
    \caption{ProbSelect vs FedLim performance evaluation.}
    \label{fig:probselect_evaluation}
\end{figure}
\subsubsection{Experiment results}
Figure~\ref{fig:slo_compliance} shows that ProbSelect consistently outperforms FedLim in meeting predefined deadline requirements. From a theoretical perspective, this translates to superior true positive rates (selected devices that meet deadlines) and lower false positive rates (devices selected but miss deadlines). ProbSelect demonstrates significant improvements across all workloads: 15.68\% higher precision for MobileNetV2, 18.75\% higher precision for ResNet-50, and 6.87\% higher precision for AlexNet. On average, ProbSelect achieves 13.77\% higher SLO compliance across all workloads, demonstrating the effectiveness of our probabilistic modeling approach. Additionally, ProbSelect demonstrates improved resource efficiency compared to FedLim. Figure \ref{fig:waste_reduction} shows that ProbSelect achieves relative waste reductions of 72.5\% for MobileNetV2, 62.7\% for ResNet-50, and 17.7\% for AlexNet, demonstrating that probabilistic client selection effectively prevents devices from consuming resources without contributing to the global model.

\section{Related Work}
\label{section:related}

Nishio and Yonetani~\cite{8761315} pioneered intelligent client selection by formulating an optimization problem that maximizes participants completing training within predefined deadlines. Building on this foundation, LEARN~\cite{9685077} extends the model to consider waiting times in TDD wireless networks, while GREED~\cite{10001569} addresses energy constraints by optimizing the trade-off between participant count and battery consumption. However, these approaches rely on CPU-centric models that fail to capture the unique computational patterns and resource requirements of GPU-accelerated training. Oort~\cite{lai2021oortefficientfederatedlearning} represents the current state-of-the-art in heterogeneous client selection, jointly optimizing statistical and system utility through an exploration-exploitation strategy. While Oort effectively balances data quality and device performance, it addresses complementary aspects of federated learning compared to our work: Oort optimizes for time-to-accuracy without deadline guarantees, while ProbSelect ensures SLO compliance through probabilistic modeling of GPU-specific training latencies. 

Adaptive deadline-based approaches have also emerged to address the client selection problem through runtime feedback mechanisms. SmartPC~\cite{9052206} proposes a hierarchical pace control framework that uses analytical models for completion time estimation combined with dynamic deadline assignment based on runtime feedback. Although SmartPC effectively balances training time and energy efficiency, it depends on offline profiling, continuous monitoring, and historical data collection, introducing operational overhead and limiting applicability in dynamic environments.



\section{Conclusion and Future Work}
\label{section:conclusion}

In this paper, we addressed the client selection problem for GPU-accelerated federated learning in the 3D continuum. We developed ProbSelect, a probabilistic approach combining analytical modeling with SLO-aware client selection. We 
proposed ALM for accurate GPU training latency estimation with MAPE below 5\%. Our evaluation demonstrates that ProbSelect effectively mitigates the straggler problem by improving SLO compliance by 13.77\% while reducing computational waste by up to 72.5\% and eliminating the need for historical data collection. Future research will explore FLOP efficiency distributions across diverse GPU architectures and workload combinations to establish statistically validated parameters.

\bibliographystyle{plain} 
\bibliography{references}

\end{document}